\documentclass[aps,prl,reprint,amsmath,amssymb,showpacs,nobibnotes]{revtex4-1}
\usepackage{graphicx}
\usepackage{hyperref} 
\usepackage{dsfont}    
\usepackage{color}

\newcommand\trick[1]{}
  
\newcommand{\be}{\begin{equation}}
\newcommand{\ee}{\end{equation}}
\newcommand{\eq}[1]{(\ref{#1})}

\newcommand{\bit}{\begin{itemize}}  \newcommand{\eit}{\end{itemize}}
\newcommand{\ben}{\begin{enumerate}}  \newcommand{\een}{\end{enumerate}}

\newcommand{\rf}[1]{(\ref{#1})}

\def\bd{\begin{document}}
\def\ed{\end{document}}
\def\bea{\begin{eqnarray}}
\def\eea{\end{eqnarray}}
\let\bm=\bibitem

\def\la{\langle}
\def\ra{\rangle}

\def\npb#1#2#3{Nucl. Phys. {\bf{B#1}} #3 (#2)}
\def\plb#1#2#3{Phys. Lett. {\bf{#1B}} #3 (#2)}
\def\prl#1#2#3{Phys. Rev. Lett. {\bf{#1}} #3 (#2)}
\def\prd#1#2#3{Phys. Rev. {D bf{#1}} #3 (#2)}
\def\cmp#1#2#3{Comm. Math. Phys. {\bf{#1}} #3 (#2)}
\def\cqg#1#2#3{Class. Quantum Grav. {\bf{#1}} #3 (#2)}
\def\nppsa#1#2#3{Nucl. Phys. B (Proc. Suppl.) {\bf{#1A}}#3 (#2)}
\def\ap#1#2#3{Ann. of Phys. {\bf{#1}} #3 (#2)}
\def\ijmp#1#2#3{Int. J. Mod. Phys. {\bf{A#1}} #3 (#2)}
\def\rmp#1#2#3{Rev. Mod. Phys. {\bf{#1}} #3 (#2)}
\def\mpla#1#2#3{Mod. Phys. Lett. {\bf A#1} #3 (#2)}
\def\jhep#1#2#3{J. High Energy Phys. {\bf #1} #3 (#2)}
\def\atmp#1#2#3{Adv. Theor. Math. Phys. {\bf #1} #3 (#2)}

\def\N{{\cal N}}
\def\sst{\scriptscriptstyle}
\def\thetabar{\bar\theta}
\def\Tr{{\rm Tr}}
\def\one{\mbox{1 \kern-.59em {\rm l}}}

\def\a{\alpha}      \def\da{{\dot\alpha}}  \def\dA{{\dot A}}
\def\b{\beta}       \def\db{{\dot\beta}}
\def\g{\gamma}  \def\G{\Gamma}  \def\dc{{\dot\gamma}}
\def\d{\delta}  \def\D{\Delta}  \def\ddt{\dot\delta}
\def\e{\epsilon}
\def\ve{\varepsilon}
\def\uve{\upvarepsilon}
\def\f{\phi}    \def\F{\Phi}    \def\vvf{\f}
\def\vphi{\varphi}
\def\h{\eta}
\def\k{\kappa}
\def\l{\lambda} \def\L{\Lambda}
\def\m{\mu} \def\n{\nu}
\def\o{\omega}
\def\p{\pi} \def\P{\Pi}
\def\r{\rho}
\def\s{\sigma}  \def\S{\Sigma}
\def\t{\tau}
\def\th{\theta} \def\Th{\Theta} \def\vth{\vartheta}
\def\X{\Xeta}
\def\z{\zeta}

\def\na{\nabla}

\def\cA{{\cal A}} \def\cB{{\cal B}} \def\cC{{\cal C}}
\def\cD{{\cal D}} \def\cE{{\cal E}} \def\cF{{\cal F}}
\def\cG{{\cal G}} \def\cH{{\cal H}} \def\cI{{\cal I}}
\def\cJ{{\mathscr J}} \def\cK{{\cal K}} \def\cL{{\cal L}}
\def\cM{{\cal M}} \def\cN{{\cal N}} \def\cO{{\cal O}}
\def\cP{{\cal P}} \def\cQ{{\cal Q}} \def\cR{{\cal R}}
\def\cS{{\cal S}} \def\cT{{\cal T}} \def\cU{{\cal U}}
\def\cV{{\cal V}} \def\cW{{\cal W}} \def\cX{{\cal X}}
\def\cY{{\cal Y}} \def\cZ{{\cal Z}}

\def\ua{\underline{\alpha}}
\def\uc{\underline{\phantom{\alpha}}\!\!\!\gamma}
\def\um{\underline{\mu}}
\def\ud{\underline\delta}
\def\ue{\underline\epsilon}
\def\una{\underline a}\def\unA{\underline A}
\def\unb{\underline b}\def\unB{\underline B}
\def\unc{\underline c}\def\unC{\underline C}
\def\und{\underline d}\def\unD{\underline D}
\def\une{\underline e}\def\unE{\underline E}
\def\unf{\underline{\phantom{e}}\!\!\!\! f}\def\unF{\underline F}
\def\unm{\underline m}\def\unM{{\underline M}}
\def\unn{\underline n}\def\unN{{\underline N}}
\def\unp{\underline{\phantom{a}}\!\!\! p}\def\unP{\underline P}
\def\unq{\underline{\phantom{a}}\!\!\! q}
\def\unQ{\underline{\phantom{A}}\!\!\!\! Q}
\def\unH{\underline{H}}

\def\As {{A \hspace{-6.4pt} \slash}\;}
\def\bs {{b \hspace{-6.4pt} \slash}\;}
\def\Ds {{D \hspace{-6.4pt} \slash}\;}
\def\Gts {{\Gt \hspace{-6.4pt} \slash}\;}
\def\ds {{\del \hspace{-6.4pt} \slash}\;}
\def\ss {{\s \hspace{-6.4pt} \slash}\;}
\def\ks {{ k \hspace{-6.4pt} \slash}\;}
\def\ps {{p \hspace{-6.4pt} \slash}\;}
\def\xs {{x \hspace{-6.4pt} \slash}\;}
\def\pas {{{p_1} \hspace{-6.4pt} \slash}\;}
\def\pbs {{{p_2} \hspace{-6.4pt} \slash}\;}
\def\cFs {{{\cal F} \hspace{-6.4pt} \slash}\;}
\def\Dss {{D \hspace{-7.5pt} \slash}\;}
\def\dss {{\del \hspace{-7.0pt} \slash}\;}

\def\Ah{{\hat{A}}}
\def\Dh{{\hat{D}}}
\def\Gh{{\hat{G}}}
\def\Fh{{\hat{F}}}
\def\Ih{{\hat{I}}}
\def\Jh{{\hat{J}}}
\def\Kh{{\hat{K}}}
\def\Lh{{\hat{L}}}
\def\Ph{{\hat{P}}}
\def\Rh{{\hat{R}}}
\def\Vh{{\hat{V}}}
\def\Xh{{\hat{X}}}

\def\ah{{\hat{\a}}}
\def\bh{{\hat{\b}}}
\def\gh{{\hat{\g}}}
\def\dh{{\hat{\d}}}
\def\rh{{\hat{\r}}}
\def\hh{\hat{h}}
\def\uh{\hat{u}}
\def\xh{\hat{x}}
\def\yh{\hat{y}}
\def\ph{\hat{p}}
\def\xih{\hat{\xi}}
\def\chih{\hat{\chi}}
\def\Psih{\hat{\Psi}}
\def\phih{\hat{\phi}}

\def\psit{\tilde{\psi}}
\def\Psit{\tilde{\Psi}}
\def\Psibt{\tilde{\bar{Psi}}}

\def\lambdat{\tilde {\lambda}}
\def\st{\tilde{\sigma}}

\def\delt{\tilde{\delta}}
\def\Phit{\tilde{\Phi}}
\def\Phitb{\overline{\tilde{Phi}}}
\def\tht{\tilde{\th}}
\def\lt{\tilde{\l}}
\def\chit{\tilde{\chi}}
\def\phit{\tilde{\phi}}

\def\At{\tilde{A}}
\def\Bt{\tilde{B}}
\def\Ct{\tilde{C}}
\def\Dt{\tilde{D}}
\def\Et{\tilde{E}}
\def\Ft{\tilde{F}}
\def\Gt{\tilde{G}}
\def\Ht{\tilde{H}}
\def\It{\tilde{I}}
\def\Jt{\tilde{J}}
\def\Qt{\tilde{Q}}
\def\Rt{\tilde{R}}
\def\Mt{\tilde{M }}
\def\Nt{\tilde{N}}
\def\St{\tilde{S}}
\def\Vt{\tilde{V}}
\def\Xt{\tilde{X}}
\def\at{\tilde{a}}
\def\ct{\tilde{c}}
\def\dt{\tilde{d}}
\def\htt{\tilde{h}}
\def\ft{\tilde{f}}
\def\gt{\tilde{g}}
\def\pt{\tilde{p}}
\def\qt{\tilde{q}}
\def\vt{\tilde{v}}
\def\nt{\tilde{n}}
\def\ut{\tilde{u}}
\def\wt{\tilde{w}}
\def\zt{\tilde{z}}
\def\xt{\tilde{x}}
\def\yt{\tilde{y}}
\def\Psit{\tilde{\Psi}}
\def\vphit{\tilde{\varphi}}
\def\tD{\tilde{\D}}

\def\eb{\bar{\epsilon}}
\def\delb{\bar{\partial}}
\def\thb{\bar{\theta}}
\def\mub{\bar{\mu}}
\def\lamb{\bar{\l}}
\def\psib{\bar{\psi}}
\def\sb{\bar{\sigma}}
\def\xib{\bar{\xi}}
\def\chib{\bar{\chi}}

\def\Psib{\bar{\Psi}}
\def\Phib{\bar{\Phi}}
\def\Lamb{\bar{\Lambda}}
\def\Sb{{\overline \Sigma}}
\def\cb{\bar{c}}
\def\hb{\bar{h}}
\def\qb{\bar{q}}
\def\wb{\bar{w}}
\def\ub{\bar{u}}
\def\zb{{\bar{z}}}
\def\Hb{\bar{H}}
\def\Qb{{\bar Q}}
\def\Omegab{\overline{\Omega}}
\def\ob{\overline{\omega}}

\def\Ab{{\overline A}} \def\Bb{{\overline B}} \def\Cb{{\overline C}}
\def\Db{{\overline D}} \def\Eb{{\overline E}} \def\Fb{{\overline F}}
\def\Gb{{\overline G}}
\def\Ib{{\overline I}}
\def\Jb{{\overline J}} \def\Kb{{\overline K}} \def\Lb{{\overline L}}
\def\Mb{{\overline M}} \def\Nb{{\overline N}} \def\Ob{{\overline O}}
\def\Pb{{\overline P}}  \def\Rb{{\overline R}}
 \def\Tb{{\overline T}} \def\Ub{{\overline U}}
\def\Vb{{\overline V}} \def\Wb{{\overline W}} \def\Xb{{\overline X}}
\def\Yb{{\overline Y}} \def\Zb{{\overline Z}}

\def\fb{{\overline f}}
\def\gb{{\overline g}}
\def\nb{{\overline n}}
\def\mb{{\overline m}}
\def\lb{{\overline l}}
\def\yb{{\overline y}}

\def\ldel{{\overleftarrow{\del}}}
\def\rdel{{\overrightarrow{\del}}}
\def\ldeldel{{\overleftarrow{\del^2}}}
\def\rdeldel{{\overrightarrow{\del^2}}}
\def\ldelb{{\overleftarrow{\bar{\del}}}}
\def\rdelb{{\overrightarrow{\bar{\del}}}}

\def\ba{{\bf a}}
\def\bk{{\bf k}}
\def\bl{{\bf l}}
\def\bp{{\bf p}}
\def\bq{{\bf q}}
\def\br{{\bf r}}
\def\bt{{\bf t}}
\def\bu{{\bf u}}
\def\bv{{\bf v}}
\def\bx{{\bf x}}
\def\by{{\bf y}}
\def\bA{{\bf A}}
\def\bR{{\bf R}}
\def\bV{{\bf V}}

\def\bz{{\boldsymbol{\zeta}}}

\def\bone{{\bf 1}}

\def\va{{\vec a}}
\def\vk{{\vec k}}
\def\vp{{\vec p}}
\def\vq{{\vec q}}
\def\vx{{\vec x}}
\def\vy{{\vec y}}
\def\vu{{\vec u}}
\def\vv{{\vec v}}
\def \vH{{\vec H}}
\def \vg{{\vec g}}

\def\vs{{\vec \sigma}}
\def\vtau{{\vec \tau}}

\newcommand{\ov}[1]{\overrightarrow{#1}}

\def\frA{\mathfrak{A}}
\def\frB{\mathfrak{B}}
\def\frC{\mathfrak{C}}
\def\frD{\mathfrak{D}}
\def\frE{\mathfrak{E}}
\def\frF{\mathfrak{F}}
\def\frG{\mathfrak{G}}
\def\frH{\mathfrak{H}}
\def\frM{\mathfrak{M}}
\def\frN{\mathfrak{N}}
\def\frR{\mathfrak{R}}
\def\frW{\mathfrak{W}}

\def\fra{\mathfrak{a}}
\def\frb{\mathfrak{b}}
\def\frf{\mathfrak{f}}
\def\frg{\mathfrak{g}}
\def\frh{\mathfrak{h}}
\def\frl{\mathfrak{l}}
\def\frs{\mathfrak{s}}
\def\fri{\mathfrak{i}}
\def\frj{\mathfrak{j}}

\def\ma{\mathfrak{a}}
\def\mg{\mathfrak{g}}
\def\mh{\mathfrak{h}}
\def\mR{\mathfrak{R}}
\def\mN{\mathfrak{N}}

\newcommand{\nn}{{\nonumber}}

\def\d{\delta}\def\D{\Delta}\def\ddt{\dot\delta}

\def\pa{\partial} \def\del{\partial}
\def\xx{\times}
\def\uno{\mbox{1 \kern-.59em {\rm l}}}

\def\trp{^{\top}}
\def\inv{^{-1}}
\def\dag{\dagger}
\def\pr{^{\prime}}

\def\rar{\rightarrow}
\def\lar{\leftarrow}
\def\lrar{\leftrightarrow}

\newcommand{\0}{\,\!}      
\def\one{1\!\!1\,\,}
\def\im{\imath}
\def\jm{\jmath}

\newcommand{\tr}{\mbox{tr}}
\newcommand{\slsh}[1]{/ \!\!\!\! #1}

\newcommand{\1}{\mbox{1}\hspace{-0.25em}\mbox{l}}

\def\vac{|0\rangle}
\def\lvac{\langle 0|}

\def\hlf{\frac{1}{2}}
\def\ove#1{\frac{1}{#1}}
\newcommand{\hot}[1]{\frac{#1}{2}}

\def\Box{\square}
\def\CC {\mathbb{C}}
\def\FF {\mathbb{F}}
\def\RR{\mathbb{R}}
\def\NN{\mathbb{N}}
\def\ZZ{\mathbb{Z}}
\def\bb#1{{\bf #1}}
\def\bcomment#1{}
\def\bfhat#1{{\bf \hat{#1}}}
\def\VEV#1{\left\langle #1\right\rangle}

\newcommand{\ex}[1]{{\rm e}^{#1}} \def\ii{{\rm i}}

\newcommand{\lrbrk}[1]{\left(#1\right)}
\newcommand{\lrsbrk}[1]{\left[#1\right]}
\newcommand{\sfrac}[2]{{\textstyle\frac{#1}{#2}}}

\def\stw{{\sqrt{2}}}

\def\rf {{\rm f}}
\def\ri {{\rm i}}
\def\rj {{\rm j}}
\def\rn {{\rm n}}
\def\rk {{\rm k}}
\def\rl {{\rm l}}
\def\rr {{\rm r}}
\def\rs {{\scriptscriptstyle \rm S}}
\def\rt {{\scriptscriptstyle \rm T}}

\def\rQ {{\scriptscriptstyle \rm \cQ}}
\def\rR {{\scriptscriptstyle \rm \cR}}

\def\cQb{{\cal \Qb}}
\def\cRb{{\cal \Rb}}
\def\cWb{{\cal \Wb}}

\def\fd {{\rm N}}
\def\afd {{\overline{\rm N}}}

\def \II {I\hspace{-.1em}I\hspace{.1em}}
\def \IIA {\mbox{\II A\hspace{.2em}}}
\def \IIB {\mbox{\II B\hspace{.2em}}}
\def \gs {g^s}
\def \ls {\lambda^s}

\def \I {{\cal I}}
\def \qs {q\hspace{-.53em}/\hspace{.15em}}
\def \ks {k\hspace{-.53em}/\hspace{.15em}}
\def \YM {{\mbox{\tiny YM}}}
\def \gym {g_{\YM}}

\def \Lc {\L_c}
\def\IR{\relax{\rm I\kern-.18em R}}
\def \id {{\bf 1}}

\def\cci{\ell}
\def\ccj{\ell'}

\def\bbq{\pmb{q}}
\def\bom{\pmb{\o}}
\def\bJ{\pmb{J}}
\def\bM{\pmb{M}}
\def\bB{\pmb{B}}
\def\bn{\pmb{n}}
\def\bE{\pmb{E}}

\newcommand{\rrr}[1]{\vskip 0.2cm \noindent{\bf #1} ---}

\begin{document}
\title{ Fermi Condensation induced by Weyl Anomaly}
\author{Chong-Sun Chu${}^{2,3}$}
\author{Rong-Xin Miao${}^{1}$ }
\affiliation{ ${}^1$ School of Physics and Astronomy, Sun Yat-Sen University,
  Zhuhai, 519082, China
  \\
${}^2$ National Center for Theoretical Sciences, National Tsing-Hua
  University, Hsinchu 30013, Taiwan\\
${}^3$ Department of Physics, National Tsing-Hua
 University, Hsinchu 30013, Taiwan
}


\date{April 14, 2020}

\begin{abstract}

Fermi condensation is usually a phenomena of strongly correlated system.
In this letter,  we point out a novel mechanism for condensation of
Dirac fermions
due to Weyl anomaly.
The condensation has its physical origin in  the nontrivial response of
  the fermion vacuum  to changes in the background spacetime
  (either boundary location or the background metric), 
  and can be felt when a background scalar field is turned on.
  The scalar field can be, for example, the Higgs field in a fundamental
  theory or the phonon in condensed matter system.
For a spacetime with boundaries, the induced Fermi condensate is
  inversely proportional to the proper distance from the boundary.
  For a conformally flat spacetime without boundaries, Fermi condensation
  depends on the conformal factor and its derivatives.
We also generalize the Banks-Casher relation which relates the Fermi condensate
to the zero mode density of the Dirac operator
to a local form.
 Due to its universal nature,  this anomaly induced Fermi condensate
 can be expected to have a wide range of applications in physics.
   
\end{abstract}

\maketitle


\rrr{1. Introduction}
Fermi condensation $\la \psib \psi \ra \neq 0$
is an interesting quantum
phenomena
and has a wide range of applications. The Cooper pair in BCS
theory of superconductivity is a famous example of Fermi
condensation. It is the bound state of a pair of electrons in a metal
with opposite spins. The chiral condensate of massless fermions is
another example of Fermi condensation. In QCD the chiral condensate is
an order parameter of transitions between different phases of quark
matter in massless limit. The condensation of Fermionic atoms has been
observed in experiment \cite{fermicondensation}.
The condensation of fermion
is usually attributed to the effects of strongly coupled dynamics and
hence it can be used as an order parameter characterizing the phases of the
theory.
A motivation of this work is to investigate if there is any
novel and universal mechanism other than strongly coupled dynamics
that can give rises to Fermi condensation.

Recently a novel phenomena of induced current was predicted in
the quantum field theory
of Dirac fermions coupled to external electromagnetic field
\be \label{S}
S=\int_M \sqrt{-g} \big(\bar{\psi} i \gamma^{\mu}\nabla_{\mu}\psi+ \bar{\psi}
  \g^\m A_\m  \psi \big).
  \ee
It was found that an applied magnetic field will
give rise to a non-uniform magnetization density of the vacuum and
induces a magnetization current
\begin{eqnarray}\label{typeIcurrent}
 \la J^{\mu}\ra = \frac{-2\beta F^{\mu\nu}n_{\nu}}{x} + \cdots, \ x\sim 0
\end{eqnarray}
in the vicinity of the boundary
of the vacuum system \cite{Chu:2018ksb,Chu:2018ntx}.
Here $\beta$ is the beta function, $x$ is the proper distance to the boundary,
$n_{\mu}$
is the inner normal vector
and $...$ denote higher order terms in
 $O(x)$. Note that the universal results (\ref{typeIcurrent})
works for
 general quantum field theory and not just conformal field
 theory. 
 In a  conformally flat spacetime $ds^2=
e^{2\sigma}\eta_{\mu\nu}dx^{\mu}dx^{\nu}$ without boundaries, the
 anomalous current is given by \cite{Chernodub:2016lbo,Chernodub:2017jcp}
  \begin{eqnarray}\label{typeIIcurrent}
\la  J^{\mu} \ra =-2\beta F^{ \mu\nu}  \partial_{\nu} \sigma+O(\sigma^2),
\end{eqnarray}
  to the leading order of small $\s$.
  Generalization of the result  \eq{typeIcurrent} to higher dimensions
  and the result  (\ref{typeIIcurrent})
  for arbitrary finite $\s$ can be found in \cite{higher1,higher2} and
  \cite{Zheng:2019xeu}
  respectively.
We remark that these
anomalous currents
do not rely on the presence of a material
system, but is a pure vacuum phenomena. This is completely different
from the other well known transport phenomena
\cite{Vilenkin:1995um,Vilenkin:1980fu,
  Giovannini:1997eg,alekseev,Fukushima:2012vr,Kharzeev:2007tn,Son:2009tf,
  Landsteiner:2011cp}
that is
due to chiral anomaly
and a finite chemical potential is needed.
In fact, 
the anomalous current \eq{typeIcurrent}
can be regarded as a kind of  magnetic Casimir effect since it arises
from the nontrivial electromagnetic
 response of the vacuum 
 to the change of boundary (location). This is very similar to
 the Casimir effect which originated from the mechanical
 response of the vacuum to the change of the boundary (location). As for
 \eq{typeIIcurrent}, it arises from the fact that the vacuum of the theory
 is different for different $\s$ and a non-vanishing vev for the
 current operator
 is resulted due to nontrivial Bogoliubov transformation.
 This is similar to
 the process of particle creation during cosmological expansion
 or the Hawking radiation \cite{davies}.

 Motivated  by these results for the anomalous current, it is natural
 to consider
 other couplings of the fermion to external field, and one can
 expect
 similar induced phenomena to occur. 
 In this paper,
 we show that, in addition to currents, Weyl anomaly
 can give rises
 to Fermi condensation if a background scalar
 field
 is turned on.
 The resulting Fermi condensate is a nontrivial function of space.
 In the standard situation of a Dirac operator coupled to external
 vector field in
 a flat spacetime, the Fermi condensate is translational invariant and
 it is well known to be related to the density of zero modes of the Dirac
 operator by the Bank-Casher relation \cite{bank}. 
 For more general Dirac operator,
 we show that the Fermi condensate obeys an elegant generalized
 form of the Bank-Casher relation, see
\eq{bc1} and
 \eq{bc2} below.

\rrr{2. Weyl Anomaly \& Fermi Condensate}
Let us start with the action of Dirac fermion $\psi$
coupled to a scalar
field $\phi$ in a curved spacetime with metric $g_{\m\n}$:
\begin{eqnarray} \label{action}
  S=\int_M \sqrt{-g} \big(i\bar{\psi}\gamma^{\mu}\nabla_{\mu}\psi+
  \phi \bar{\psi}
    \psi \big).
\end{eqnarray}
In this letter, we use the signature $(1,-1,-1,-1)$ for the metric
and 
the scalar field will be taken as a background and so
whether there is a Lagrangian 
for the scalar field
is irrelevant to us.
In this theory, the
renormalized expectation value of the Fermi condensate
$\la \bar{\psi} \psi \ra$
can be
derived by the variation of effective action with respect to the
background scalar field
\begin{eqnarray} \label{Fermicondensation}
  \la \bar{\psi} \psi \ra
  =\frac{1}{\sqrt{-g}}\frac{\delta I_{\rm eff}}{\delta \phi}.
\end{eqnarray}
The action \eq{action} is classically Weyl invariant under the local scaling
transformation: $\psi \to e^{ -\s} \psi$, $g_{\m\n} \to e^{2\s} g_{\m\n}$
and $ \phi \to e^{- \s} \phi.$
However, quantum mechanically there is an anomaly.
Imposing the bag boundary conditions \cite{Chodos:1974je,Chodos:1974pn}  $(1\pm \gamma_5 \gamma_n)\Psi|_{\partial M}=0$ and applying the heat kernel
expansion \cite{Vassilevich:2003xt}, we obtain the Weyl anomaly
$\cA (g_{\m\n}, \phi)$
at one loop as
\begin{eqnarray} \label{anomaly}
  \mathcal{A}=
  \frac{1}{8\pi^2}\left(\int_M \sqrt{-g} \big[-(\nabla \phi)^2+
    \frac{R\phi^2}{6}+\phi^4\big] \right. \nn\\
  + \left. \int_{\partial M}\sqrt{-h} \frac{k \phi^2}{3} \right).
\end{eqnarray}
Here $R$ is Ricci scalar in the bulk $M$, $h_{ij}$ is the induced
metric on the boundary $\partial M$,
$k_{ij}$ is the extrinsic curvature and $k$ is its trace. 
Note that we have ignored the gravitational
contribution to Weyl anomaly, i.e, $\int_M O(R^2)+\int_{\partial M}
O(Rk)$, since they are independent of $\phi$ and hence are irrelevant for the Fermi condensation.
In the following, 
we  show 
that the knowledge of the Weyl
anomaly (\ref{anomaly}) allows one to determine
the Fermi condensate
in closed analytic form
without performing any perturbative QFT calculation.
This is one of the main results
of this letter.

\rrr{3. Fermi Condensation I: Boundary Theory}
Let us first study the Fermi condensation
in 4-dimensional spacetime with
a boundary, say, at $x=0$ of a certain coordinate system.
We follow the methods of
\cite{Miao:2017aba,Chu:2018ksb}, where we have studied the expectation
value of current and stress tensor in boundary quantum field theories
\cite{Cardy:2004hm}.  As the mass dimension of $\psi$ is 3/2,
the Fermi condensate takes
the asymptotic form \cite{Deutsch:1978sc}
\begin{eqnarray} \label{FermicondensationI0}
\la \bar{\psi}
\psi \ra =\frac{O_0}{x^3}+\frac{O_1}{x^2}+\frac{O_2}{x}+O(\ln x)
\end{eqnarray}
near the boundary. Here $x$ is the proper distance from the boundary.
$O_\rn$ have mass dimension $\rn$ and depend on only the background
geometry and the background scalar field.
For example, we have $O_0=c_0$, $O_1=c_1\phi+c_2 k$, 
 where $c_i$ are numbers.
Since Weyl anomaly is related to the UV Logarithmic
divergent term of effective action, one can follow
the same analysis performed in \cite{Miao:2017aba,Chu:2018ksb}
and obtain the following
``integrability'' relation \cite{sign}
\begin{eqnarray} \label{key}
(\delta \mathcal{A})_{\partial M}=\left(\int_M \sqrt{-g} \la \bar{\psi}
  \psi \ra \delta \phi \right)_{\log\epsilon}
\end{eqnarray}
between the
renormalized
Fermi condensate and the boundary part of the variation
of the Weyl anomaly.
Here a regulator $x \geq \e$ to the boundary has been introduced for the
integral on RHS of \eq{key}. For our purpose, we turn  only on the
variation of the scalar field. Using (\ref{key}), one can
derive Fermi condensate near the boundary from the boundary terms of
variations of the Weyl anomaly. To proceed, let us
employ the Gauss normal coordinates to write the metric
$ds^2=dx^2+ (h_{ij}-2x k_{ij}+
\cdots )dy^i dy^j $ and expand 
$\phi(x)=\phi_0+ x
\phi_1+ x^2 \phi_2+O(x^3)$, where $x\in [0,+\infty)$ and $\phi_i$
  give the $i^{th}$ derivatives of $\phi$ at $x=0$.
  From (\ref{anomaly}), we get
  for
  the LHS of
  (\ref{key})
\begin{eqnarray} \label{keyLHS}
(\delta \mathcal{A})_{\partial M}=\frac{-1}{4\pi^2}\int_{\partial
    M}\sqrt{-h}(\nabla_{n}\phi -\frac{1}{3}k\phi)\delta \phi_0,
\end{eqnarray}
where $n$ denotes the internal normal of the manifold $M$.
 Next, we substitute
 (\ref{FermicondensationI0})
into the RHS of (\ref{key}),
integrate over $x$
and select the logarithmic divergent term, we obtain
\begin{eqnarray} \label{keyRHS}
  \int_{\partial M}\sqrt{-h}\big[ -O_0 \delta \phi_2 + (k O_0-O_1) \delta
  \phi_1-(O_2+\cdots) \delta \phi_0 \big],\; \;\;\;\;\;\;
\end{eqnarray}
where $\cdots$ denotes terms which depend on $O_0$ and $O_1$
which vanish
when $O_0=O_1=0$.  Comparing (\ref{keyLHS}) and (\ref{keyRHS}), we obtain
$O_0=O_1=0$ and  the  Fermi condensation near the boundary:
\begin{eqnarray}
  \label{FermicondensationI1}
\la \bar{\psi} \psi \ra= \frac{1}{4\pi^2}\frac{\nabla_{n}\phi
  -\frac{1}{3}k\phi}{x} +O(\ln x), \quad x \sim 0.
\end{eqnarray}

Several comments are in order.
{\tt 1}. Similar to the case of current and stress tensor
\cite{Chu:2018ksb,Miao:2017aba}, the Fermi condensate is finite at the boundary
since there are boundary contributions to
the Fermi condensation which cancel the divergence from the bulk contribution
(\ref{FermicondensationI1}).
{\tt 2}.
The result (\ref{FermicondensationI1}) is for the bag
boundary conditions at one loop. 
For general boundary conditions, there are corrections to Weyl
anomaly (\ref{anomaly}) and
$O_0$ and $O_1$ of (\ref{FermicondensationI0}) are non-zero \cite{Chu:2020gwq}. 
{\tt 3}. 
The result
(\ref{FermicondensationI1})
for the bag boundary condition, as well as
the general form (\ref{FermicondensationI0}) for general boundary
conditions can be reproduced in holographic theory \cite{Chu:2020gwq}.
{\tt 4}. Similar to
\cite{Miao:2017aba,Chu:2018ksb}, the above discussions on
Weyl-anomaly-induced Fermi condensation apply  not only to conformal
field theory (CFT) but also the general quantum field theory (QFT). That is
because Weyl anomaly is well-defined for general quantum field
theories \cite{Duff:1993wm,Brown:1976wc}. {\tt 5}. If $\phi$ is the Higgs field
and get  a vev 
$\phi =-m$, 
then the fermion get a mass $m$ and the
chiral symmetry is spontaneous broken. Our analysis implies that
a Fermi condensate is induced near a curved boundary  
\begin{eqnarray} \label{FermicondensationI2}
\la \bar{\psi} \psi\ra= \frac{1}{12\pi^2}\frac{ m k }{x}, \quad x \sim 0
\end{eqnarray}
due to the Higgs phenomena. 
{\tt 6}. Note that the above discussions apply to more general theories instead of just the free theory (\ref{action}). In fact, there are universal relations between the Fermi condensate and Weyl anomaly  \cite{Chu:2020gwq}.

\begin{figure}[t]
\centering
\includegraphics[width=6cm]{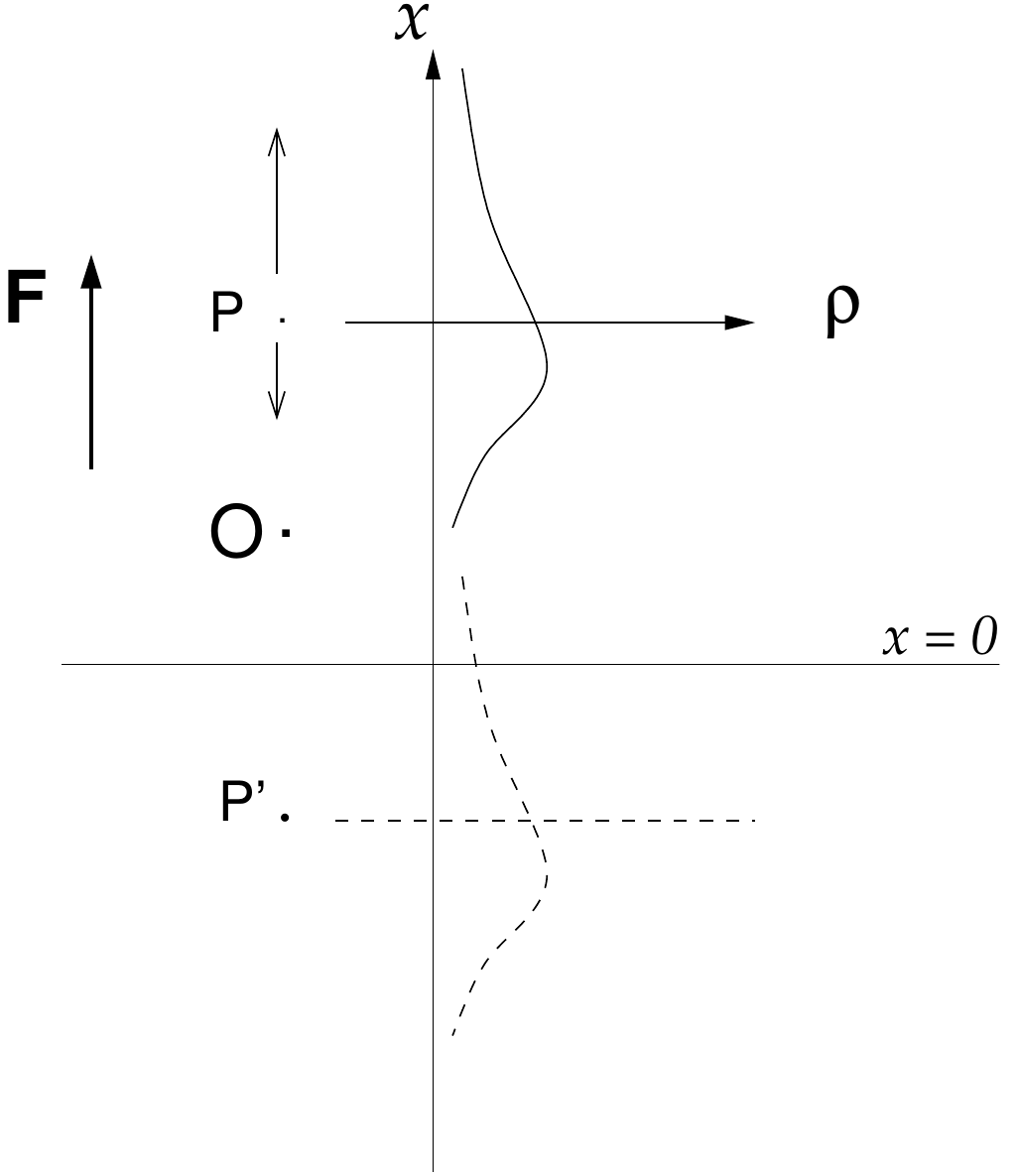}
\caption{Fermi condensate near the boundary. The contribution from $P'$
is absent when half space $x>0$ is considered.}
\label{f1}
\end{figure}

Similar to the origin for the induced current discussed in
\cite{Chu:2018ksb,Chu:2018ntx},
the existence of the Fermi condensation \eq{FermicondensationI1}
can be understood in terms of the changed
properties of the vacuum due to the presence of boundary.
For simplicity let us consider the case of
a half space flat geometry $x \geq 0$.
From the form of the action \eq{action}, the fermions see a potential
of $-\phi$ and experience a
force
\be \label{force}
F_x = \del_x \phi
\ee
in the $x$-direction. This may also be derived
from the Ehrenfest theorem
$
\la \frac{d \cO}{dt} \ra = \frac{1}{i \hbar} \la [\cO, H] \ra
$
for the time evolution of the observable $\cO$. For the Hamiltonian
$H =- \int d^3 x \psib (i \g^i \del_i +\phi) \psi$, we obtain the
the ``Newton's law''
$ \la \frac{dp_x}{dt} \ra = \del_x \phi \la \psib \psi  \ra $
for the $x$-direction, and $\la dp_a / dt\ra =0$ for $a= y,z$.
This gives the force  \eq{force}
on single particle states.
For simplicity, let us consider the case of a constant force $\del_x \phi>0$.
Consider the pair creation process at any point $P$ in space. Due to the presence
of an upward force (for the discussion here, we will call the positive $x$ direction
as upward), the particles
which are created to move upward will, for the same period of time $\Delta t$,
travel a longer distance compared to  
the particles which are created to move downward. Note that the force does not
differentiate particles and anti-particles and so the
number density $\rho_P$ for the fermions and anti-fermions
contributed by source point $P$ will be affected by the force the same way.
As a result, $\rho_P$
will not be symmetric with respect to $P$, but is
skewed towards the
negative $x$ direction. See figure \ref{f1}. Now take any observation
point $O$ in space. The amount of condensate at $O$ is obtaining by summing over
the contribution from all source points $P$. When there is no boundary,
obviously a constant condensate $\la \psib \psi \ra$ is created. Things are
different when there is a boundary at $x=0$. For a source point $P$ situated near
the boundary, the particles
that are created to move downward will have
less chance to be annihilated since the space $x<0$ are all
absent now and hence there are less particles that can
travel up to annihilate them. As a result, the distribution of
particles will be skewed even more towards the boundary. Also the skewing
is greater as one gets closer to the boundary. 
Summing up the contributions from all source points,
one see immediately that the condensate is bigger towards the boundary.
Also it is positive and is proportional to the force $\del_x \phi$.
Obviously the same
discussion holds for the case of $\del_x \phi<0$. Qualitatively,
we find exact agreement with the
result \eq{FermicondensationI1}.
It is worth noting that since the scalar force is the same for the fermions and the
anti-fermions, the virtual pairs are pushed towards the boundary in the same
manner and there is no separation of charges.
Charge neutrality of the vacuum is maintained. 

It is interesting to remark that the condensate \eq{FermicondensationI1} of
fermions near the boundary can also be 
understood as a kind of Bose-Einstein condensation with localization in
coordinate space.
Heuristically, due to the force \eq{force}, the virtual particles experience an
acceleration $ a = F_x/m$ and hence an Unruh temperature 
\cite{Unruh:1976db}
\be
T = |a|/(2\pi k).
\ee
For the discussion here, we will allow for both signs of $\del_x \phi$.  
Since the condensate is a bosonic state and obeys the Bose-Einstein statistics,
the mean occupation number of the condensate is given by
$ N = 1/(e^{E/kT} -1)$. Now the particles created at a location $x$ has a potential
energy of $E \sim F_x x$. Near the boundary we have $E\ll kT$ and
$N \sim kT/E \sim \text{sgn}(\del_x \phi)/mx$.
The condensate $\la \psib \psi \ra$ has mass dimension 3
and should depend on $m$ and $\del_x \phi$. The natural relation 
$\la \psib \psi \ra \sim m |\del_x \phi| \times N$ then gives precisely the result
\eq{FermicondensationI1}. 


\rrr{4. Fermi Condensation II: Conformally Flat Spacetime}
Fermi condensation can also occur in 
conformally flat spacetime without boundaries.
To demonstrate this, let us start by deriving the anomalous
transformation rule for the Fermi condensate. 
Consider the theory \eq{action} with metric and scalar field given by
$(g_{\m\n},\phi)$. Due to the anomaly, the renormalized effective
action $I_{\rm eff}$ is not invariant under the Weyl transformation.
Generally, we have \cite{footnote1}
$\frac{\d}{\d\s} I_{\rm eff}(e^{-2\s} g_{\m\n}, e^{\s}\phi) =
\cA (e^{-2\s} g_{\m\n}, e^{\s}\phi)$
for arbitrary finite $\s(x)$.
This can be integrated to give the effective
action \cite{Wess:1971yu,Cappelli:1988vw,Schwimmer:2010za}. Using the fact that
the anomaly (\ref{anomaly}) is Weyl invariant up to a surface term:
$ \cA(e^{-2\s} g_{\m\n}, e^{\s} \phi)
= \cA(g_{\m\n}, \phi) + \int_M
\del_\m(\sqrt{-g} \phi^2 g^{\m\n} \del_\n \s),$
we obtain immediately the transformation rule for the effective action:
\bea\label{Ieff-t}
&& I_{\rm eff}(e^{-2\s} g_{\m\n}, e^{\s}\phi)=  I_{\rm eff}(g_{\m\n}, \phi)  \\
&& +\frac{1}{8\pi^2}\int_M \sqrt{-g}
\left[\left( -(\nabla \phi)^2+\frac{R\phi^2}{6}
    +\phi^4\right) \sigma+\frac{\phi^2}{2}(\nabla \sigma)^2\right], \nn
\eea
plus a boundary term
$\frac{1}{24\pi^2}\int_{\partial M}\sqrt{-h} k \phi^2 \sigma$,
which we drop in spacetime without boundaries. One can
check that the dilaton effective action satisfies Wess-Zumino consistency
$[\delta_{\sigma_1},\delta_{\sigma_2}] I_{\rm eff}=0$.  Using \eq{Ieff-t},
we obtain finally the transformation rule for the Fermi
condensate \eq{Fermicondensation} under Weyl transformation
$g_{\mu\nu} \to g'_{\mu\nu} = e^{-2\sigma} g_{\m\n}$, $\phi \to \phi' = e^\s \phi$,
\bea \label{FermicondensationI3}
\la\bar{\psi}
\psi\ra = - \frac{1}{4\pi^2}\Big[
   \nabla(\sigma\nabla\phi)+(2\phi^3+\frac{1}{6}\phi
    R) \sigma+\frac{1}{2}\phi (\nabla\sigma)^2\Big], \nn\\
  \eea
  plus a trivial term $e^{-3\sigma}\la\bar{\psi} \psi\ra' $.
  Here $\la\bar{\psi} \psi\ra$ (resp.  $\la\bar{\psi} \psi\ra'$)
  denotes the vev of the Fermi condensate of
  the theory \eq{action} in the background spacetime $g_{\mu\n}$
  (resp. $g'_{\mu\n}$).
Taking $g'_{\mu\nu}$ to be the flat spacetime metric and the fact that the
Fermi condensation vanishes in flat spacetime,
we finally obtain \eq{FermicondensationI3} as the
Fermi condensate in conformally flat spacetime 
\be\label{cf-metric}
ds^2 = e^{2 \s} \eta_{\m\n}dx^\m dx^\n.
\ee

Several comments are in order.  {\tt 1}. The conformal factor $\sigma$
of (\ref{FermicondensationI3}) is arbitrary and needs not to be
small. As a result, we can use (\ref{FermicondensationI3}) to
calculate Fermi condensation in general conformally flat spacetimes
such as Anti-de-Sitter space, de-Sitter space and Robertson-Walker
universe. {\tt 2}. For Robertson-Walker universe
$ds^2=dt^2-a(t)^2(dx^2+dy^2+dz^2)$, 
we have at time $t=t_*$
\begin{eqnarray} \label{FermicondensationI4}
\la\bar{\psi} \psi\ra= - \frac{1}{4\pi^2}(H\dot{\phi}+\frac{1}{2}H^2\phi)
\end{eqnarray}
where dot denotes time derivative and $H=\dot{a}/a$ is the Hubble
parameter. For simplicity we have chosen $a(t_*)=1$.  {\tt 3}. Unlike
the case with boundaries, (\ref{FermicondensationI3}) works well only
for CFT.  
At high energy scale such as early
universe, the particle mass can be ignored and fermions can be
regarded as 
CFT
approximately. Then all of the
above discussions apply.  {\tt 4}. The result
(\ref{FermicondensationI3}) can also be derived
for strongly coupled conformal field
theory that is dual to gravity \cite{Chu:2020gwq}.
 {\tt 5}. 
The physical reason for the condensate \eq{FermicondensationI3}
is simple.
It arises because the vacuum
of the theory in the spacetime \eq{cf-metric} at a conformal factor $\s$ is no
longer a vacuum as the metric changes to have a different conformal
factor $\s'$.
As a result, a fermion condensate is created
in a way similar to
the creation of particles due to cosmological expansion
(see for example, \cite{davies}). 

\rrr{5. Generalized Banks-Casher Relation}
The Banks-Casher relation links the spontaneous breaking of chiral symmetry in
QCD to the presence of a non-zero density of zero modes of the
Dirac operator of the quark field. In ordinary consideration where the QFT is
translational invariant, the condensate is
a constant and one has the following relation \cite{bank}
\be \label{bc-std}
\la \psib \psi \ra = \pi \r(0),
\ee
where $\r(\l)$ is the spectral density of the Dirac operator,
see \eq{rho} below.
Here let us generalize the
Banks--Casher relation to the case where the condensate is not a constant.
Consider
a theory of Dirac fermions in curved space coupled to an external field $X$:
\be
S = \int_M \sqrt{-g} i \psib D \psi, \quad
\mbox{where} \quad iD := i \g^\m \nabla_\m +X. 
\ee
For example, $X= \g^\m A_\m$ gives the minimal
coupling to  gauge potential $A_\m$, $X= \phi$ gives the Yukawa coupling to
a scalar field, and $X= i \g^{\m\n} F_{\m\n}$ gives the Pauli coupling to
gauge field
strength. Let $\psi_n$ be the eigenstate of
the Dirac operator
$iD$ with eigenvalue $\l_n$,
and satisfies the orthogonormal relation
$ \int_M \sqrt{-g} \psi_n^\dag \psi_m = \d_{nm}$. 
It follows from the path integral definition of chiral condensate 
$\la \psib \psi \ra = Z^{-1} \int \cD\psi \cD\psib e^{-S} \psib\psi$
that
\be \label{bc-non-int-1}
\la \psib \psi \ra = - \sum_n \psi^\dag_n(x) \frac{1}{\lambda_n}\psi_n(x).
\ee
It is convenient to introduce the spectral density
\be \label{rho}
\r (\l) := \frac{1}{V} \sum_n \d(\l -\l_n),
\ee
where $V$ is the spacetime volume.
The relation \eq{bc-non-int-1} takes the form
\be  \label{bc-non-int-2}
\frac{i}{V}\la \psib \psi \ra =
-\int_{-\infty}^\infty d\l \r(\l) \frac{1}{\l +i \e} \psi^\dag_\l(x) \psi_\l(x),  
\ee
where a Wick rotation
$\psib \to i \psib$ on the spinor has been performed to obtain the
relation \eq{bc-non-int-2} in Minkowski space. We have introduced
an $i \e$ prescription ($\e>0$) to regulate the divergence due to the
existence of
zero modes of the operator $iD$.  Using the relation
$\frac{1}{\l + i\e} = {\rm P}\, \frac{1}{\l} - i \pi \d(\l),$
where P denote the Cauchy Principal value, we obtain 
the following {\it generalized Banks-Cashier relation} in the un-integrated
form:
\be \label{bc1}
\frac{1}{V} \la \psib \psi (x) \ra
= \pi \rho(0) |\psi_0(x)|^2 + i{\rm P} \int_{-\infty}^\infty d\l \rho(\l)
\frac{1}{\l}  |\psi_\l(x)|^2.
\ee
This can also be integrated to give
\be \label{bc2}
\frac{1}{V} \int_M \sqrt{-g} \la \psib \psi \ra = \pi \r(0) +
i {\rm P} \int_{-\infty}^\infty d \l \; \frac{\r(\l) }{\l}.
\ee
For flat space, the condensate is a constant due to translational invariance.
For operator which anticommute with $\g^5$:
$\{ iD, \g^5\} =0$
(e.g. minimal coupling to external gauge field),
the eigenvalues
of $iD$ always come in pairs $\pm \l$ and 
it is 
$\psi_{-\l} = \g^5 \psi_\l$ for $\l \neq 0$.
In this case, the continuum contribution
in the RHS of \eq{bc2} vanishes 
and we obtain the standard Banks-Cashier \eq{bc-std}.
For the general case of a curved spacetime
with a non-minimal coupling, the condensate
can become imaginary in general, and the condensate is related to the spectral
density as in \eq{bc1} and
\eq{bc2}. For the
case of Yukawa coupling in 
conformally flat spacetime,
the condensate is real and we obtain the constraint
\be
{\rm P} \int_{-\infty}^\infty d\l \rho(\l)
\frac{1}{\l}  |\psi_\l(x)|^2 =0,
\ee
even through the spectrum is not symmetric.

\rrr{6. Conclusions} 
In this letter we have shown that the nontrivial vacuum structure of
the theory of Dirac fermions in the presence of an external scalar field can
lead to the occurrence of Fermi condensation. 
This is a direct consequence of the violation of scaling symmetry in the
presence of a background scalar field that couples to the fermion.
In general, it is also interesting to consider the effect of
external pseudoscalar field on the Fermi condensation \cite{Chu:2020gwq}.
Due to the simple and universal nature of the Yukawa coupling,
one can expect
the Fermi condensation \eq{FermicondensationI1}, \eq{FermicondensationI3}
to find a wide range of physical applications.
For
example the possible occurrence of a condensate during the inflationary
phase of the universe may have non-trivial consequences on various
dynamical aspects
of the early Universe 
such as the reheating,
the evolution of physical structure,
the pattern of symmetry realization etc.
Experimentally, it may be possible to make observation of the
Fermi condensate in optomechanical system with boundary or in a thermal system
where temperature can
be stimulated by
a conformally flat background \cite{lut}.\\

\begin{acknowledgments}
We thank T. W. Chiu, B. L. Hu, S. Iso, S. Lin and G. Shiu
for useful discussions.        
C. S. Chu was supported  by 
NCTS and the grant MOST 107-2119-M-007-014-MY3 
of MSTT.
R. X. Miao acknowledges the NSFC grant (No. 11905297).
Note that all Institutes contribute equally to this work, the order of Institutes is adjusted for the assessment policy of SYSU.
\end{acknowledgments}






\begin{thebibliography}{99}
\bibitem{footnote1}
  We have find it conveniently to consider the Weyl transformation
  $g_{\mu\nu} \to e^{-2\s} g_{\mu\nu}$ here.
\end{thebibliography}
\end{document}